# One Dimensional Modeling Of Dielectric Barrier Discharge in Pure Oxygen at Atmospheric Pressure Using Comsol Multiphysics

Mohammed Habib Allah LAHOUEL[#1], Djilali BENYOUCEF[1], Abdelatif GADOUM [1]

[#]Technology Department, Hassiba Benbouali University
Laboratoire Génie Electrique et Energies Renouvelables, Chlef University, Algeria

[1]HABIB221@live.com

*Abstract*— **This paper present a model of dielectric barrier discharge DBD describe and simulate the electrical breakdown at atmospheric pressure in pure oxygen using Comsol Multiphysics for the better understanding and explanation of physical behaviours of the existing species in gap. For this model, we prefer to simulate using the 1D geometry using physics-based kinetic methods. DBD have a several applications, such as ozone generation, surface treatment, light source and other environmental industries. This model include a numerical and chemical model. Courant and voltage characteristics presented in this paper. Density of existing and newly created species in gap also presented.**

## I. INTRODUCTION

The main characteristic of a Dielectric Barrier Discharge DBD device lays in the setup of the electrodes. In particular, the setup includes the presence of a dielectric layer within the discharge gap distance d insulating a. The dielectric presence preventing the formation of arcs or sparks, as well as of a steady plasma state. To sustain the discharge, AC voltage or a pulsed one is applied to the electrodes at a frequency ranging from several hundreds of hertz to a few hundreds of kHz. [1]. The DBD modeling in high voltage electric power applications are mainly based on its static behavior and taking into account the charges that deposited in various amounts on the dielectric surface and calculating the electric field in the gap. Determination of plasma conditions (like gas temperature) and plasma parameters (electron distribution function and electron density) are required for effective optimization, and thereby the safe application of the plasma sources. [2]. Modeling of DBD allow for the study of basic plasma processes, but also enable analysis and optimization of current technologies using plasmas, as well as predicting the performance of as-yet-unbuilt systems for new applications.[3] Copious quantities of reactive oxygen species are often generated in atmospheric pressure plasmas. The reactive species and amounts generated depend on plasma parameters, such as the rotational, vibrational, excitation temperature and electron density.[4] this proposed model include plasma chemistry, surface reactions, density of electrons and heavy species (ions, atom…), electric failed, electron temperature, for better understanding dbds at atmospheric pressure in $O_2$.

## II. NUMERICAL MODEL

### A. Electron Transport

The continuity equation for the electron density is given by:

$$\frac{\partial n_e}{\partial t} + \nabla \cdot \Gamma_e = R_e \qquad (1)$$

$$\Gamma_e = -\mu_e E n_e - \nabla(D_e n_e) \qquad (2)$$

With:

$n_e$: Electron density $(1/m^3)$

$R_e$: Electron rate expression $(1/m^3 \cdot s)$

$\mu_e$: Electron mobility $(m^2/V \cdot s)$

$E$: Electric field $(V/s)$

$D_e$: Electron diffusivity $(m^2/s)$

$e$: refers to electron.

Electron energy is given by:

$$\frac{\partial n_\varepsilon}{\partial t} + \nabla \cdot \Gamma_\varepsilon + E \cdot \Gamma_e = R_\varepsilon \qquad (3)$$

$$\Gamma_\varepsilon = -\mu_\varepsilon E n_\varepsilon - \nabla(D_\varepsilon n_\varepsilon) \qquad (4)$$

With:

$n_\varepsilon$: Electron energy density $(V/m^3)$

$R_\varepsilon$: Energy loss/gain due to inelastic collisions $(V/(m^3 \cdot s))$

$\mu_\varepsilon$: Electron energy mobility $(m^2/V \cdot s)$

$D_\varepsilon$: Electron energy diffusivity $(m^2/s)$

$\varepsilon$: refers to energy.

Electron energy, $\varepsilon$ $(V)$ computed by:

$$\varepsilon = \frac{n_\varepsilon}{n_e} \qquad (5)$$

The source coefficients in the above equations are determined by the stoichiometry of the system and are written using either rate coefficients or Townsend coefficients.

Suppose that there are M reactions which contribute to the growth or decay of electron density and P inelastic electron-neutral collisions.

The electron energy loss is obtained by summing the collisional energy loss over all reactions:

$$R_e = \sum_{j=1}^{P} x_j k_j N_n n_e \Delta\varepsilon_j \quad (6)$$

With:

P: inelastic electron-neutral collisions;
$x_j$: Mole fraction of the target species for reaction j;
$k_j$: Rate coefficient for reaction j (m³/s);
$N_n$: Total neutral number density (1/m³);
$\Delta\varepsilon_j$: Energy loss from reaction j (V);

The Townsend coefficients $k_j$ depend exponentially on the mean electron energy, $\varepsilon$. When a Maxwellian EEDF is assumed, the Townsend coefficients can be fitted with a function of the form:

$$k_j = A\varepsilon^\beta e^{-E/\varepsilon} \quad (7)$$

B. *Electrostatic Equations*

- Electrostatic field :
$$-\nabla . \varepsilon_0 \varepsilon_r \nabla V = \rho \quad (8)$$
- The space charge density :
$$\rho = q\left(\sum_{k=1}^{N} Z_k n_k - n_e\right) \quad (9)$$
- The constitutive relation :
$$D = \varepsilon_0 \varepsilon_r E \quad (10)$$

This relation (9) describe the macroscopic properties of the dielectric coating medium (relating the electric displacement D with the electric field E) and the applicable material properties.

- Surface charge accumulation is added to boundaries by way of the following boundary condition:
$$-n.(D_1 - D_2) = \rho_s \quad (11)$$

where $\rho_s$ is the solution of the following distributed ODE on the boundary:

$$\frac{d\rho_s}{dt} = n.J_i + n.J_e \quad (12)$$

Where $n.J_i$ and $n.J_e$ are the normal component of the total ion current density and the total electron current density respectively on the wall.

### III. PLASMA CHEMISTRY

In this part of our work is an aggregation of possible chemical reactions in oxygen $O_2$ as a gas filled in discharge gap.

A. *Electron impact reactions*

The cross section set for electron collisions in $O_2$ used in this calculation is shown in TABLE I [5.6].

One requirement for data set is that it leads to calculated electron transport coefficients which are reasonably consistent with experiment. A second requirement is that the energy dependence of the cross sections be consistent with electron beam experiments [5].

TABLE I
PHELPS DATABASES

| Reaction | Formula |
|---|---|
| 1 | e+ $O_2$ =>e+ $O_2$ |
| 2 | e+ $O_2$ =>O+O⁻ |
| 3 | e+ $O_2$ =>e+O2a1d |
| 4 | e+O2a1d=>e+ $O_2$ |
| 5 | e+ $O_2$ =>e+O2b1s |
| 6 | e+O2b1s=>e+ $O_2$ |
| 7 | e+ $O_2$ =>e+O245 |
| 8 | e+O245=>e+ $O_2$ |
| 9 | e+ $O_2$ =>e+O+O |
| 10 | e+ $O_2$ =>e+O+O1d |
| 11 | e+ $O_2$ =>e+O+O1s |
| 12 | e+ $O_2$ =>e+e+ $O_2^+$ |
| 13 | e+O2a1d=>e+O2a1d |
| 14 | e+O2a1d=>e+O+O |
| 15 | e+O2a1d=>2e+ $O_2^+$ |
| 16 | e+O2b1s=>e+O2b1s |
| 17 | e+O2b1s=>e+O+O |
| 18 | e+O2b1s=>2e+ $O_2^+$ |
| 19 | e+O245=>e+O+O |
| 20 | e+O245=>2e+ $O_2^+$ |
| 21 | e+O=>e+O |
| 22 | e+O=>e+O1d |
| 23 | e+O1d=>e+O |
| 24 | e+O=>e+O1s |
| 25 | e+O1s=>e+O |
| 26 | e+O=>2e+O⁺ |
| 27 | e+O1d=>e+O1s |
| 28 | e+O1d=>2e+O⁺ |
| 29 | e+O1s=>2e+O⁺ |
| 30 | e+ $O_3$ =>e+ $O_3$ |
| 31 | e+ $O_3$ =>O⁻+ $O_2$ |
| 32 | e+ $O_3$ =>2e+ $O_3^+$ |

B. *Other chemical reactions:*

The reactions used in our model token with their rate constant $k^f$. Quantitatively the Arrhenius Equation determines the relationship between the rate constant proceeds and its temperature [7].

$$k^f = A e^{-E_a/RT} \quad (13)$$

With:
A: Pre-exponential factor;
T: Temperature (kelvin);
$E_a$ : Activation energy for the reaction (in the same units as R*T);
R: Universal gas constant [8].

C. *Surface reactions*

Besides to the previous used chemical reactions, the following surface reactions are implemented (TABLE II):

TABLE III
SURFACE REACTIONS

| Reaction | Formula |
|---|---|
| 1 | $O_2^+$ => $O_2$ |
| 2 | O⁺=>O |
| 3 | $O_3^+$ => $O_3$ |
| 4 | O2a1d=> $O_2$ |
| 5 | O2b1s=> $O_2$ |
| 6 | O245=> $O_2$ |
| 7 | O1d=>O |
| 8 | O1s=>O |
| 9 | $O_2^-$ => $O_2$ |
| 10 | O⁻=>O |

### IV. SIMUALATION RESULTS

For this model, we consider a parallel point geometry with a gap filled by oxygen.

Voltage used, is a sinusoidal function with high frequency f=50 kHz (Fig.1) as :

$$Vrf = 6000 * sin(2 * \pi * f * t) \quad (14)$$

Where t is time (μs), and Vrf applied voltage on the internal circle (V) and the external circle in related to the ground.

The dielectric considered with relative permittivity $\varepsilon_r = 10$.

- Initial values:

Number of seed electrons $10^3 (1/m^3)$ is supposed to be present in gap. Those are uniformly distributed in gap length with the initial mean energy 4 V.

Number density of oxygen ions supposed the same initial density of seed electrons, $10^3 (1/m^3)$.

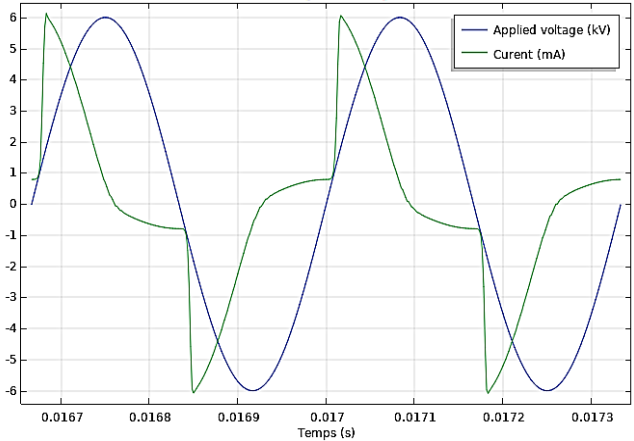

Fig.1 Applied voltage and the result total discharge courant

The Figures 2 and 3 show the electron density, ion Oxygen number density in gap respectively.

When the voltage applied, an electric field forms in the gap. Free electrons in the gap will be accelerated and may acquire enough energy to cause ionization. The newly created electrons rush towards the grounded dielectric. An equal number of ions are also created since the electrons and ions must be created in equal pairs. The ions rush towards the dielectric surface where the voltage is being applied with negative polarity. However due to be heavier than electrons, the ions are nearly the grounded dielectric in Figure 3.

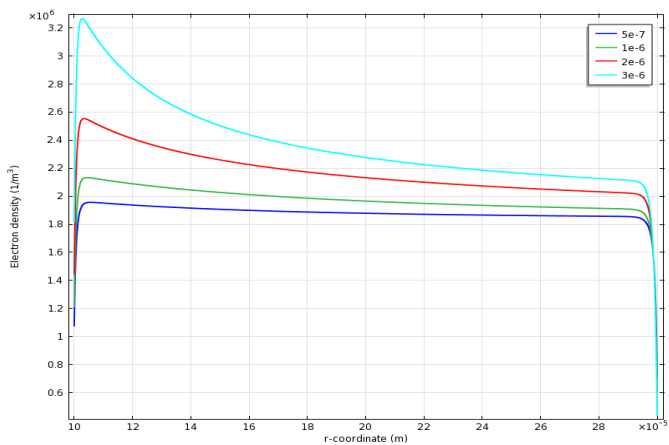

Fig.2 Electron Number density

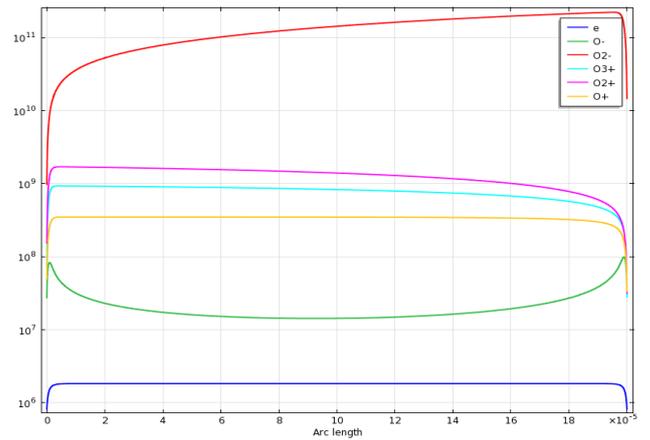

Fig.3 Ion Oxygen number density

From Figure 2 and 3, it can be understood that there is a great rise in number densities of electron and Oxygen ions the ions and free electrons concentration is near the high voltage electrode with positive polarity that means until this time avalanches occur in this area. However as the applied voltage increases, a stronger electric field forms in the gap and any free electron in the gap will be accelerated and since the electric field is strong enough they acquire enough energy to cause significant ionization activities that means avalanches occurs in across the gap uniformly.

## V. CONCLUSIONS

This paper present a model study of Oxygen discharge in atmospheric pressure with the existence of dielectric barrier based on plasma physics. Our model highlight the physics of the breakdown process by using a simple plasma chemistry model. The computational simulations were presented for applying a sinusoidal voltage it is showed:

- the creation of new elements during the discharge like Ozone $O_3$ for example;
- Most importantly the negative oxygen ions are a key element in producing plasma actuator's